\documentclass[twocolumn,aps,prl,superscriptaddress,amsmath,amssymb]{revtex4}
\usepackage{amsfonts}
\usepackage{bbm}
\usepackage{color}

\usepackage{graphicx}
\usepackage{dcolumn}
\usepackage{bm}

\begin{document}
	
\title{Resonant tunneling in disordered borophene nanoribbons with line defects}
\author{Pei-Jia Hu}
\affiliation{Hunan Key Laboratory for Super-microstructure and Ultrafast Process, School of Physics and Electronics, Central South University, Changsha 410083, China}

\author{Si-Xian Wang}
\affiliation{Hunan Key Laboratory for Super-microstructure and Ultrafast Process, School of Physics and Electronics, Central South University, Changsha 410083, China}

\author{Xiao-Feng Chen}
\affiliation{Hunan Key Laboratory for Super-microstructure and Ultrafast Process, School of Physics and Electronics, Central South University, Changsha 410083, China}
\affiliation{School of Physical Science and Technology, Lanzhou University, Lanzhou 730000, China}

\author{Zeng-Ren Liang}
\affiliation{Hunan Key Laboratory for Super-microstructure and Ultrafast Process, School of Physics and Electronics, Central South University, Changsha 410083, China}

\author{Tie-Feng Fang}
\affiliation{School of Sciences, Nantong University, Nantong 226019, China}

\author{Ai-Min Guo}
\email[]{aimin.guo@csu.edu.cn}
\affiliation{Hunan Key Laboratory for Super-microstructure and Ultrafast Process, School of Physics and Electronics, Central South University, Changsha 410083, China}

\author{Hui Xu}
\affiliation{Hunan Key Laboratory for Super-microstructure and Ultrafast Process, School of Physics and Electronics, Central South University, Changsha 410083, China}

\author{Qing-Feng Sun}
\affiliation{International Center for Quantum Materials, School of Physics, Peking University, Beijing 100871, China}
\affiliation{Collaborative Innovation Center of Quantum Matter, Beijing 100871, China}
\affiliation{CAS Center for Excellence in Topological Quantum Computation, University of Chinese Academy of Sciences, Beijing 100190, China}

\date{\today}

\begin{abstract}
Very recently, borophene has been attracting extensive and ongoing interest as the new wonder material with structural polymorphism and superior attributes, showing that the structural imperfection of line defects (LDs) occurs widely at the interface between $\nu_{1/5}$ ($\chi_3$) and $\nu_{1/6}$ ($\beta_{12}$) boron sheets. Motivated by these experiments, here we present a theoretical study of electron transport through two-terminal disordered borophene nanoribbons (BNRs) with random distribution of LDs. Our results indicate that LDs could strongly affect the electron transport properties of BNRs. In the absence of LDs, both $\nu_{1/5}$ and $\nu_{1/6}$ BNRs exhibit metallic behavior, in agreement with experiments. While in the presence of LDs, the overall electron transport ability is dramatically decreased, but some resonant peaks of conductance quantum can be found in the transmission spectrum of any disordered BNR with arbitrary arrangement of LDs. These disordered BNRs exhibit metal-insulator transition by varying nanoribbon width with tunable transmission gap in the insulating regime. Furthermore, the bond currents present fringe patterns and two evolution phenomena of resonant peaks are revealed for disordered BNRs with different widths. These results may help for understanding structure-property relationships and designing LD-based nanodevices.
\end{abstract}
\maketitle

\begin{figure}
\includegraphics[width=8.6cm]{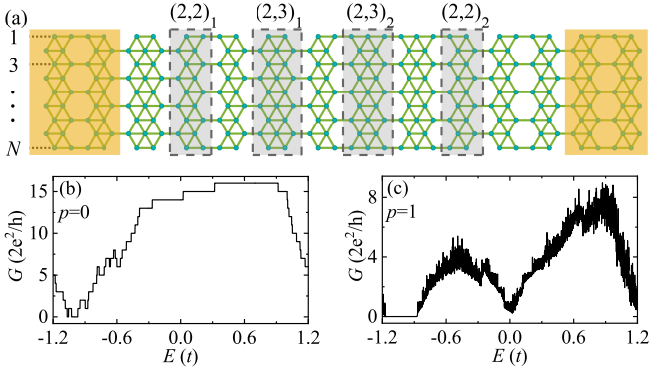}
\caption{\label{fig1} (a) Schematics of a two-terminal disordered BNR coupled to left and right semi-infinite $\nu_{1/5}$ BNRs (yellow rectangles). This disordered BNR is assembled from random arrangement of (2,2) and (2,3) chains (gray rectangles) in the CSR, where a (2,2)$_j$ chain is followed randomly by a (2,2)$_{\bar{j}}$ or (2,3)$_{\bar{j}}$ one and a (2,3)$_j$ chain by a (2,2)$_j$ or (2,3)$_j$ one, with $\bar{j}$=1 (2) for $j$=2 (1). This leads to structural disorder induced by randomly distributed LDs and the disorder is characterized by the ratio $p$ of (2,3) chains to all of the chains (length $L$). Here, the length is $L=10$ and the width defined as the number of rows is $N=9$. Energy-dependent conductance $G$ for (b) a $\nu_{1/5}$ BNR with $p=0$ and (c) a $\nu_{1/6}$ BNR with $p=1$.}
\end{figure}

$Introduction$.--Borophene, a monolayer of boron atoms, has attracted extensive attention as a prototype for exploring novel two-dimensional (2D) systems \cite{zzes,ajm2,ld1} since its successful synthesis by two independent groups \cite{ajm,bfj} following theoretical predictions \cite{th1,lkc1,yx1,th2,th3,pes1,wx1}. In contrast to other 2D materials \cite{ssd,mp1,bls}, multiple borophene polymorphs including freestanding ones have been realized experimentally \cite{qzhong,wli,rwu,bki,nav,pran,fb1}, which exhibit in-plane anisotropy and are tunable by ambient growth conditions. Of particular interest are the planar $\rm\nu_{1/5}$ and $\rm\nu_{1/6}$ boron sheets, which consist of (2,2) and (2,3) chains, respectively. Here, the indices $a_n$ and $a_w$ of ($a_n$,$a_w$) represent, respectively, the number of atoms in the narrow and wide rows of a single boron chain \cite{qzh}, as illustrated in Fig.~\ref{fig1}(a). These two sheets possess many intriguing attributes, such as excellent mechanical strength and flexibility \cite{zz1,zzh}, conventional superconductivity \cite{pes2,mga,zy1}, Dirac fermions \cite{bfo,mez,bfjs}, and ultrahigh thermal conductance \cite{ld1,zh1}.


On the other hand, line defects (LDs) exist in diverse 2D materials \cite{jla,vdz,ns1,co1,lm1,pr1} and strongly affect their electronic, magnetic, optical, mechanical, and thermal properties \cite{dgu,lkou,jts,hhu,khp1,jbd,lx1,kt1,mvk,hk1,wf1}. Very recently, a specific LD at the interface between (2,2) and (2,3) chains has been widely reported in experiments \cite{xli,lx2,ywa,ljk,lli,qli,xli2}, owing to higher structural stability of borophene in the presence of LDs \cite{ywa,sgx2}. Specifically, several distinct phases are synthesized from periodic self-assembly of LDs \cite{xli,lx2,ywa}, implying that (2,2) and (2,3) chains function as building blocks to construct novel boron sheets. These LDs could considerably modulate the electronic properties of borophene \cite{xli,ljk,jze} and improve its mechanical response \cite{sgx2}, thus playing an important role in the discovery of exotic quantum phenomena and in device applications. Notice that the LDs in realistic borophene lack long-range order and structural disorder emerges simultaneously \cite{xli,ywa}. Until now, the LD-induced structural disorder has not yet been discussed and understanding this disorder may facilitate the elucidation of structure-property relationships.

In this Letter, we study theoretically the electron transport through two-terminal borophene nanoribbons (BNRs) with random distribution of LDs by connecting to left and right semi-infinite $\nu_{1/5}$ BNRs. These BNRs, termed as disordered BNRs, are assembled from random arrangement of (2,2) and (2,3) chains in the central scattering region (CSR) and leads to LD-induced structural disorder, as shown in Fig.~\ref{fig1}(a). We find that both $\nu_{1/5}$ and $\nu_{1/6}$ BNRs exhibit metallic behavior, consistent with experiments \cite{pran,fb1,qzh,xli}. Remarkably, although the overall electron transport efficiency is dramatically declined in the presence of LDs, some resonant peaks of conductance quantum appear in the transmission spectra of various disordered BNRs, regardless of nanoribbon length and LD distribution. By changing nanoribbon width, the disordered BNRs could present metal-insulator transition with tunable transmission gap in the insulating regime. Furthermore, two evolution phenomena of resonant peaks are revealed for disordered BNRs with different widths, showing that (i) all of the resonant peaks for odd width overlap perfectly with all those of two narrow BNRs and (ii) a specific resonant peak for width $N_i$ reappears at various BNRs with width $N=\alpha(N_i +1)-1$ and $\alpha$ an integer.

{\it Model}.--Electron transport through two-terminal disordered BNRs is simulated by the Hamiltonian \cite{bfo,mez}: $\mathcal{H}=\sum_{i} \epsilon_{i} a_{i}^{\dagger} a_{i}- t \sum_{\langle i,j\rangle} a_{i}^{\dagger} a_{j}$. Here, $a_{i}^{\dagger}$ ($a_{i}$) creates (annihilates) an electron at site $i$, $\epsilon_{i}$ is the potential energy, and $t$ is the nearest-neighbor hopping integral. According to the Landauer-B\"{u}ttiker formula, the conductance is expressed as $G = G_0 {\rm Tr} [\mathbf{\Gamma} _{\rm L} \mathbf{G}^{r} \mathbf{\Gamma} _{\rm R} \mathbf{G}^{a} ]$. The conductance quantum $G_0= 2 e^2 /h $, the Green's function $\mathbf{G}^{r} (E)= [\mathbf{G} ^{a} (E) ]^ {\dagger}= [ E\mathbf{I} -\mathbf{H}_{\rm c} -\mathbf{\Sigma}_ {\rm L}^{r}- \mathbf{\Sigma} _{\rm R}^{r} ] ^{-1}$, and the linewidth function $\mathbf{\Gamma}_{\rm L/R}=i (\mathbf{\Sigma}_{\rm L/R}^{r}-\mathbf{\Sigma}_{\rm L/R}^{a} )$, with $E$ the electron energy, $\mathbf{H}_{\rm c}$ the CSR Hamiltonian, and $\mathbf{\Sigma} ^{r} _{\rm L/R}$ the retarded self-energy due to the coupling to the left/right semi-infinite $\nu_{1/5}$ BNR \cite{mpl1,mpl2}.

In the numerical calculations, the parameters are fixed to $\epsilon_i =0$, $t=1$ (energy unit), and the length counted by all of the chains in the CSR is taken as $L=2000$, unless stated otherwise. We consider the most disordered BNRs with half of the chains being (2,3) ones ($p=0.5$) and the results are averaged over 2500 disordered samples.

{\it Pure} BNRs.--We first study the electron transport through pure BNRs with the CSR being a $\nu_{1/5}$ or $\nu_{1/6}$ BNR, as shown in Figs.~\ref{fig1}(b) and~\ref{fig1}(c). It clearly appears that both $\nu_{1/5}$ and $\nu_{1/6}$ BNRs exhibit metallic behavior which is consistent with experiments \cite{pran,fb1,qzh,xli}, and their conductances are asymmetric about the line of $E=0$ owing to the electron-hole symmetry breaking. For the $\nu_{1/5}$ BNR, its transmission spectrum is characterized by many conductance plateaus quantized at integer multiples of $\rm G_{0}$ as expected [Fig.~\ref{fig1}(b)], because of the translational symmetry. By contrast, when the $\nu_{1/5}$ BNR is replaced by a $\nu_{1/6}$ one, the conductance declines and presents dramatic oscillating behavior instead of quantized plateaus [Fig.~\ref{fig1}(c)], in accordance with first-principles calculations \cite{jze}, because of LD-induced scattering at the CSR-lead interfaces.

\begin{figure}
\includegraphics[width=8.6cm]{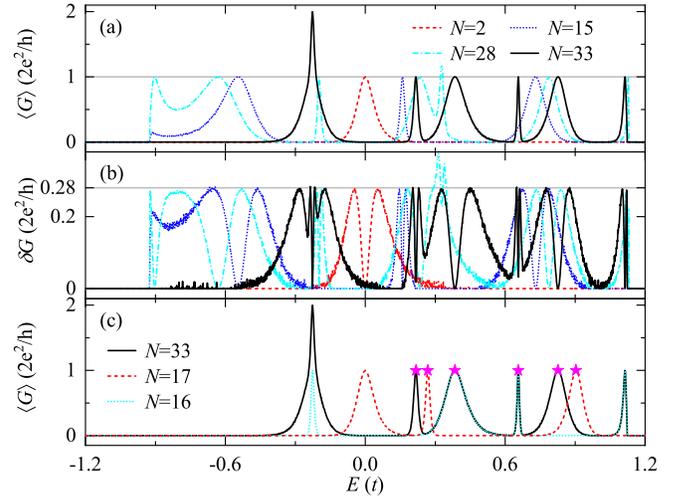}
\caption{\label{fig2} Electron transport along disordered BNRs with different widths. Energy dependence of (a) averaged conductance $\langle G \rangle$ and (b) standard deviation $\delta G$ for typical $N$. (c) $\langle G \rangle$ vs $E$ for three $N$. The gray lines in (a) and (b) denote the conductance quantum $G_0$ and $\delta G=0.28 G_0$, respectively.}
\end{figure}

{\it Disordered} BNRs.--Figures~\ref{fig2}(a) and~\ref{fig2}(b) show energy-dependent averaged conductance $\langle G \rangle$ and standard deviation $\delta G$, respectively, for disordered BNRs with different widths $N$. Here, $\delta G \equiv \sqrt{\langle G^2 \rangle -\langle G \rangle^2}$. One can see that the electron transport through disordered BNRs is strongly suppressed as expected, due to Anderson localization caused by successive scattering from randomly distributed LDs \cite{sm}. However, it is surprising that some transmission peaks of $\langle G \rangle =G_0$ are found in all these disordered BNRs and the peak number usually increases with $N$ [Figs.~\ref{fig2}(a) and~\ref{fig4}(b)]. In particular, the standard deviation at peak positions satisfies $\delta G\sim 0$ \cite{note1} and these peaks are robust against $L$ and $p$ \cite{sm}, implying that the transmission peaks could be observed in any disordered BNR with arbitrary arrangement of LDs. This result explains a recent experiment that delocalized states are measured in BNRs with LDs \cite{ljk}. These transmission peaks originate from resonant tunneling \cite{sm}, where some resonant energies remain unchanged for $\nu_{1/6}$ BNRs with different $L$ and the others change with $L$. Therefore, the electrons with invariant resonant energies cannot be reflected by LDs, as further demonstrated from spatial distributions of bond currents \cite{sm}, leading to resonant peaks in disordered BNRs.

Besides, one can identify other important features. (i) The transport property of disordered BNRs strongly depends on $N$, ranging from metallic [see the red-dashed line in Fig.~\ref{fig2}(a)] to insulating behavior [see the other lines in Fig.~\ref{fig2}(a)]. This indicates the width-driven metal-insulator transition in disordered BNRs. And the transmission gap is tunable in the insulating regime by varying $N$, which is similar to graphene nanoribbons \cite{em,syw,hmy} and facilitates band-gap engineering of BNRs. (ii) The resonant peaks, characterized by full width at half maximum (FWHM), are mainly categorized into narrow and wide peaks separated by about $0.06t$ FWHM [Figs.~\ref{fig2}(a) and~\ref{fig4}(b)]. (iii) Each resonant conductance peak corresponds to two peaks in the curve of $\delta G$-$E$ [Fig.~\ref{fig2}(b)]. Remarkably, the maximum of all these peaks satisfies $\delta G \sim 0.28G_0$ which is close to $\sqrt{1/12} G_0$ reported in disordered graphene nanoribbons \cite{ada,jcce}. (iv) Some transmission peaks can even achieve $2G_0$ [see the black-solid line in Fig.~\ref{fig2}(a)], because of almost perfect superposition of two neighboring resonant peaks, as seen from the four-peak structure around $E\sim -0.226t$ in Fig.~\ref{fig2}(b).

{\it Evolution phenomenon (EP)} I.--We then focus on the evolution of resonant peaks in disordered BNRs with various $N$. Figure~\ref{fig2}(c) shows $\langle G \rangle$ vs $E$ for disordered BNRs with three $N$. It clearly appears that some resonant peaks for $N=33$ are superimposed on all those for $N=16$, while the remaining ones will overlap with all those for $N=17$ by properly moving peak positions. This phenomenon can also be observed in other disordered BNRs with, e.g., $N=5$, $11$, $13$, and $27$ \cite{sm}. Therefore, we conclude that all of the resonant peaks of disordered BNRs with odd width $N_o$ could be assembled from the ones with $N=(N_o-1)/2$ and $(N_o+1)/2$, namely, EP I.

\begin{figure}
\includegraphics[width=8.6cm]{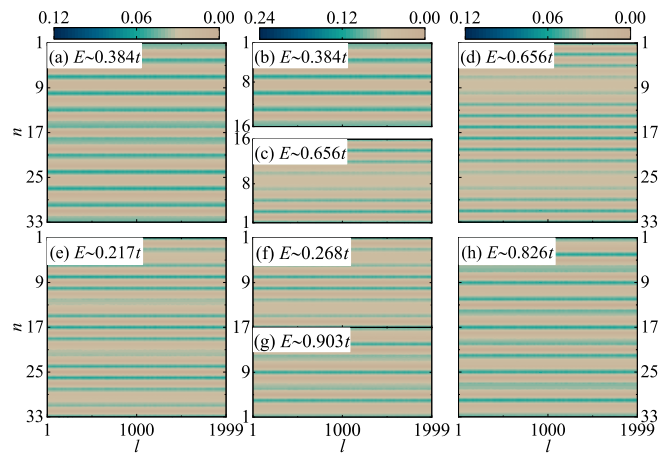}
\caption{\label{fig3} Spatial distributions of averaged interchain currents $\langle I_{l}(n) \rangle$ of disordered BNRs at the resonant energies marked by the stars in Fig.~\ref{fig2}(c). 2D plots of $\langle I_{l}(n) \rangle$ vs chain index $l$ and row index $n$ for $N=33$ at (a) $E\sim 0.384t$, (d) $E\sim 0.656t$, (e) $E\sim 0.217t$, and (h) $E\sim 0.826t$, for $N=16$ at (b) $E\sim 0.384t$ and (c) $E\sim 0.656t$, and for $N=17$ at (f) $E\sim 0.268t$ and (g) $E\sim 0.903t$.}
\end{figure}

To elucidate the underlying physics of EP I, Fig.~\ref{fig3} plots spatial distributions of averaged interchain currents $\langle I_l(n) \rangle$ at typical resonant energies shown in Fig.~\ref{fig2}(c). The currents flowing from the $l$th chain to the $l+1$th one read \cite{pjh,acre,chle}: $\mathbf{I}_{l} =(-2e/h) \int [\mathbf{H}_ {l,l+1} \mathbf{G} _{l+1,l} ^{<}(\xi) -\mathbf{H}_ {l+1,l} \mathbf{G} _{l,l+1} ^{<} (\xi)] d\xi$, where $\mathbf{G}_{l,l+1}^{<}$ is the interchain lesser Green's function. It is clear that all of the $\langle I_l(n) \rangle$ exhibit rather uniform fringe patterns and are independent of $l$ albeit the existence of randomly distributed LDs, manifesting delocalized states in various disordered BNRs. And green fringes denote transmission channels with finite $\langle I_l(n) \rangle$, whose number and location depend on $E$.

Since the spatial inversion symmetry with respect to the row of $n=(N+1)/2$ is preserved for disordered BNRs with odd $N$ [Fig.~\ref{fig1}(a)], the corresponding $\langle I_l(n) \rangle$ possess mirror symmetry [Figs.~\ref{fig3}(a) and \ref{fig3}(d)-\ref{fig3}(h)] and can be divided into two cases according to the parity of wave functions. (i) The $\langle I_l(n) \rangle$ at $n=17$ are zero for $N=33$ [Figs.~\ref{fig3}(a) and \ref{fig3}(d)], corresponding to odd wave functions whose electron densities are zero along the symmetric row of $n=17$. In this situation, the top segment from $n=1$ to 16 and the bottom one from  $n=18$ to 33, separated by this symmetric row, are independent of each other. Consequently, the transmission channels of the top/bottom segment for $N=33$ are the same as those for $N=16$ [Figs.~\ref{fig3}(a) and \ref{fig3}(b); Figs.~\ref{fig3}(c) and \ref{fig3}(d)], giving rise to completely identical resonant peaks at the same $E$ of these disordered BNRs. (ii) Contrarily, the $\langle I_l(n) \rangle$ are finite at $n=17$ [Figs.~\ref{fig3}(e) and \ref{fig3}(h)], which refers to even wave functions with nonzero electron densities along the symmetric row. Interestingly, the transmission channels of the top/bottom segment for $N=33$ are also identical to those for $N=17$ albeit distinct magnitude of $\langle I_l(n) \rangle$ at the $n$th row [Figs.~\ref{fig3}(e) and \ref{fig3}(f); Figs.~\ref{fig3}(g) and \ref{fig3}(h)]. Correspondingly, the disordered BNRs with $N=17$ and 33 possess identical resonant peaks but at different resonant energies, owing to the interference effect at the symmetric row. By contrast, the spatial inversion symmetry is destroyed for disordered BNRs with even $N$ and the parity of wave functions disappears simultaneously, leading to the absence of EP I for even $N$.

\begin{figure}
\includegraphics[width=8.6cm]{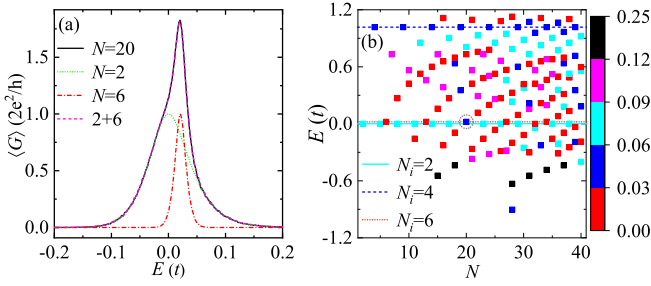}
\caption{\label{fig4} Evolution of resonant peaks in disordered BNRs with different widths. (a) $\langle G \rangle$ vs $E$ around a peak for three $N$. The magenta-dashed line represents the sum of $\langle G \rangle$ for $N=2$ and 6. (b) A 2D plot of FWHM for all of the peaks vs $N$ and $E$. The cyan-solid, blue-dashed, and red-dotted lines display the evolution of three peaks initially appeared in disordered BNRs with width $N_i=2$, 4, and 6, respectively.}
\end{figure}


{\it EP} II.--Figure~\ref{fig4}(a) shows $\langle G\rangle$ vs $E$ around a transmission peak for three disordered BNRs, where the magenta-dashed line represents the sum of $\langle G \rangle$ for $N=2$ and 6. One can see that the peak at $E\sim 0.021t$ for $N=20$ overlaps perfectly with the superposition of two neighboring peaks for $N=2$ and 6, as seen from the black-solid and magenta-dashed lines in Fig.~\ref{fig4}(a). This indicates that a transmission peak of wide disordered BNRs could be evolved from narrow ones.


To further elucidate the above phenomenon, Fig.~\ref{fig4}(b) displays a 2D plot of FWHM for all of the peaks as functions of $N$ and $E$. It is clear that the number of squares increases with $N$ and some disordered BNRs possess peculiar resonant peaks with FWHM exceeding $0.12t$ [see the leftmost peak for $N=15$ in Fig.~\ref{fig2}(a) and the black squares in Fig.~\ref{fig4}(b)]. In particular, completely identical resonant peaks are observed at $E\sim 0$ for $N=2$, 5, 8, ..., at $E\sim 1.017t$ for $N=4$, 9, 14, ..., and at $E\sim 0.021t$ for $N=6$, 13, ... [see the cyan-solid, blue-dashed, and red-dotted lines in Fig.~\ref{fig4}(b)]. Therefore, we infer that a resonant peak, firstly emerged in a disordered BNR with $N=N_i$, will reappear at the same $E$ of various BNRs with $N= \alpha(N_i +1)-1$, where $\alpha$ is an integer and $N_i +1$ the period. This characteristic is named as EP II, which compensates EP I. Notice that the peak at $E\sim 0.021t$ for $N=3 \times7-1=20$ deviates from EP II [see the surrounded blue square in Fig.~\ref{fig4}(b)], because of finite-size effects. By increasing $L$, this transmission peak is split into two peaks at $E\sim 0$ and $0.021t$ [Fig.~\ref{fig4}(a)]. Consequently, this surrounded square will be evolved into a cyan (red) square on the cyan-solid (red-dashed) line and EP II holds.

To gain insights into EP II, we consider a disordered BNR with $N=\alpha (N_i + 1)-1$. According to its unique structure, this disordered sample could be divided into $\alpha$ basic BNRs with width $N_i$, where the $m$th BNR includes the rows from $n=m_i-N_i$ to $m_i-1$ and is separated from the $m+1$th one by the $m_i$th row, with $0< m < \alpha$ and $m_i= m(N_i+1)$. For instance, the BNR with $N=9$ can be divided into two basic BNRs with $N_i=4$, which are separated by the fifth row [Fig.~\ref{fig1}(a)]. The Hamiltonian $\mathbf{H}_c$ of this disordered BNR with width $N$ can then be partitioned as:
\begin{eqnarray}
\begin{aligned}
\mathbf{H}_c=\left[\begin{array}{ccccc}
\mathbf{H}_1 & \mathbf{A}_{11} & \mathbf{0} & \cdots & \mathbf{0} \\
\mathbf{A}_{11}^\dagger & \mathbf{R}_1 & \mathbf{A}_{21} & \ddots & \vdots \\
\mathbf{0} & \mathbf{A}_{21}^\dagger & \mathbf{H}_2 & \ddots & \mathbf{0} \\
\vdots & \ddots & \ddots & \ddots &  \mathbf{A}_{\alpha, \alpha-1} \\
\mathbf{0} &\cdots & \mathbf{0} & \mathbf{A}_{\alpha, \alpha-1}^\dagger & \mathbf{H}_\alpha \\
\end{array}\right],\label{eq1}
\end{aligned}
\end{eqnarray}
where $\mathbf{H}_m$ and $\mathbf{R}_m$ are the sub-Hamiltonians of the $m$th basic BNR and the $m_i$th row, respectively, and $\mathbf{A}_{mn}$ the hopping matrix from the $m_i$th row to the $m$th BNR when $n=m$ and from the $m$th BNR to the  $m_i-1$th row when $n=m-1$. Since the $m$th and $m+1$th BNRs are mirror images about the $m_i$th row, the eigenstates of $\mathbf{H}_m$ and $\mathbf{H}_ {m+1}$ are the same with identical eigenenergies, and the hopping matrices satisfy $\mathbf{A} _{m+1,m}= \mathbf{A}_ {m,m}^\dagger$. Assuming the resonant state of the $m$th basic BNR is described by the Schr\"{o}dinger equation of $\mathbf{H}_m| \mathbf{ \Phi}_0 \rangle= E_r| \mathbf{\Phi}_0 \rangle$, with $| \mathbf{ \Phi}_0\rangle$ the wave function and $E_r$ the resonant energy, the wave function $|\mathbf{\Psi} \rangle$ of $\mathbf{H}_c$ can then be constructed as:
\begin{eqnarray}
|\mathbf{\Psi} \rangle= \sqrt{\alpha} /\alpha (| \mathbf{ \Phi}_0\rangle, \mathbf{0}, -| \mathbf{\Phi} _0\rangle, ..., (-1)^{ \alpha -1} | \mathbf{\Phi} _0\rangle)^{\rm T}.\label{eq2}
\end{eqnarray}
Substituting Eqs.~(\ref{eq1}) and (\ref{eq2}) into the Schr\"{o}dinger equation, one obtains $\mathbf{H}_c| \mathbf{ \Psi} \rangle= E_r| \mathbf{\Psi} \rangle$ straightforwardly. Thus, disordered BNRs with $N=\alpha (N_i+1)-1$ possess a completely identical resonant peak for different $\alpha$. These results are further confirmed by the numerical results of bond currents \cite{sm}.

$Conclusion$.--In summary, the electron transport along disordered BNRs with random distribution of LDs is investigated. Our results indicate that although the overall electron transport efficiency is strongly declined, some resonant peaks are observed in the transmission spectra and leads to delocalized states in various disordered BNRs. These disordered BNRs could exhibit metal-insulator transition by varying nanoribbon width with tunable transmission gap in the insulating regime. Besides, the evolution of resonant peaks in disordered BNRs with different widths is explored, which is related to the spatial inversion symmetry.

Pei-Jia Hu thanks Jing-Ting Ding for useful discussions. This work is supported by the National Natural Science Foundation of China (Grants No. 11874428, No. 11874187, and No. 11921005), the National Key Research and Development Program of China (Grant No. 2017YFA0303301), and the High Performance Computing Center of Central South University.


\begin{references}
\bibitem{note1} The standard deviation is nonzero at the resonant energies (i.e., peak positions) of $E \sim 0.329t$ for $N=28$ and of $E \sim 0.217t$ and $0.656t$ for $N=33$, because the conductance contributed only from neighboring resonant peak(s) at these energies fluctuates due to finite-size effects-induced long tails of transmission peaks. We emphasize that $\delta G \sim 0$ at these resonant energies for longer disordered BNRs.
\end{references}
\end{document}